\documentclass[a4paper,twocolumn,fleqn]{article}

\usepackage[english]{babel}
\usepackage{newtxtext}
\usepackage{mathtools}
\usepackage{amsthm}
\usepackage{newtxmath}
\usepackage{graphicx}
\usepackage{cite}

\usepackage{bm}
\usepackage{siunitx}
\usepackage{xcolor}
\usepackage{soul}
\usepackage{placeins}
\usepackage{float}

\usepackage{pfr_preprint_2026}

\usepackage{hyperref}

\newcommand{\AN}[1]{#1}

\newcommand{\rev}[1]{#1}
\newcommand{\marker}[1]{#1}

\begin{document}

\title{Two-component Bayesian modeling in the data analysis of \\
collective Thomson scattering}


\author{Akito~NAKANO\sup{1,2}, Kentaro~SAKAI\sup{1,2}, Hantao~JI\sup{3,4}, Lan~GAO\sup{4}, Brandon~K.~RUSSELL\sup{3}, and Takeo~HOSHI\sup{1,2}}

\affiliation{
  \sup{1}Department of Fusion Science, The Graduate University of Advanced Studies (SOKENDAI), 322-6 Oroshicho, Toki, Gifu 509-5292, Japan \\
  \sup{2}National Institute for Fusion Science, 322-6 Oroshicho, Toki, Gifu 509-5292, Japan\\
  \sup{3}Department of Astrophysical Sciences, Princeton University, Princeton, New Jersey 08544, USA\\
  \sup{4}Princeton Plasma Physics Laboratory, Princeton University, Princeton, New Jersey 08543, USA}

\email{nakano.akito@nifs.ac.jp}


\begin{abstract}

We apply Bayesian inference to the analysis of collective Thomson scattering (CTS) spectra, a diagnostic technique for determining local plasma parameters. Using the population annealing Monte Carlo (PAMC) method, we evaluate the posterior probability density functions (PDFs) for six plasma parameters. The analysis is performed separately for the ion acoustic wave (IAW) and electron plasma wave (EPW) features and jointly for both features. The results show that the combined analysis provides constraints on all six parameters, although the degree of constraint depends on the parameter, whereas the separate analyses provide strong constraints only on complementary subsets of the parameters. In particular, the combined analysis provides a clear constraint on the ion charge state, which is not determined by either separate analysis alone. These findings demonstrate the importance of Bayesian modeling that accounts for the two-component nature of plasmas, consisting of ions and electrons.


\end{abstract}



\maketitle  

\section{Introduction}
\label{sec:1}
Bayesian inference plays a crucial role in the analysis of experimental data, particularly when the available data are limited \AN{\cite{bishop2006prml,vontoussaint2011bayesian}.
Bayesian inference requires an appropriate probabilistic model for the measurement process; this construction is called Bayesian modeling \cite{fischer2003bayesian}.
Such a model should be designed on the basis of prior knowledge of the experimental method \cite{fischer2003bayesian}.
For plasma diagnostics, Bayesian modeling should also account for the fact that plasmas consist primarily of two components: ions and electrons \cite{froula11}.}

This paper focuses on collective Thomson scattering (CTS), a unique diagnostic tool for measuring local plasma parameters in high-energy-density plasmas \cite{froula11,milder21prl,tomita23srep}. CTS is a regime of Thomson scattering in which the Debye length is shorter than the probe laser wavelength and is described by the parametric resonance among incident and scattered electromagnetic waves and electrostatic fluctuations in plasmas: electron plasma waves (EPWs) and ion acoustic waves (IAWs) in unmagnetized plasmas. Whereas non-collective Thomson scattering, which corresponds to Doppler spectroscopy of electrons, reflects the electron velocity distribution function, CTS can provide information on both electron and ion velocity distribution functions through resonance with EPWs and IAWs. Typically, for unmagnetized plasmas with Maxwellian velocity distribution functions, CTS spectra exhibit four resonant peaks: two associated with EPWs and two with IAWs. Due to the large difference in the phase velocities of EPWs and IAWs, the required wavelength ranges are different for EPW and IAW features, and these spectra are measured with different spectrometers; one is a broadband spectrometer for the EPW resonant peaks, and the other a high-resolution spectrometer for the IAW resonant peaks. 

By fitting observed CTS spectra with the model \cite{froula11}, one can extract velocity distribution functions or plasma parameters.
Even for data obtained from the same experiment, discrepancies can arise between the parameters inferred from the EPW and IAW features. These discrepancies stem from various factors, including spectral noise characteristics, instrumental response functions, and plasma inhomogeneities. Conventionally, discrepancies are explained by introducing gradients of physical quantities within the measurement volume, and the resulting gradient-aware interpretations are supported by comparison with hydrodynamic simulations or atomic models \cite{follett16rsi}. \marker{However, the reliance on human intervention and judgment in this process can introduce subjective} bias. This raises concerns regarding the transparency and objectivity of data interpretation when inferring physical quantities based solely on observations. 
Bayesian inference is an effective method for reducing bias in data analysis and has recently been applied to CTS \cite{foo23aipa,escalona23srep}, mainly to estimate key parameters and associated uncertainties using low-dimensional posterior \AN{probability density functions (PDFs)} or error-bar evaluations. Although these works demonstrated the usefulness of Bayesian inference for CTS analysis, they did not address the full high-dimensional structure of the \AN{posterior PDF} or the simultaneous treatment of both EPW and IAW features.

\marker{Here, we propose a Bayesian model based on prior knowledge of the electron (EPW) and ion (IAW) features.}
The CTS analysis with this tailored Bayesian model is carried out using the data-analysis framework ODAT-SE (formerly known as 2DMAT) \cite{motoyama22cpc}. 
ODAT-SE is a general-purpose data-analysis framework and has been applied to data-analysis problems in plasma science \cite{kobayashi26rsi,sakai26ppcf,sakai26jinst,kin26pfr}, as well as in materials science \cite{hoshi22cpc,sato25prm,liu25prl}.
The remainder of the present paper is organized as follows. Section \ref{SEC-METHOD} explains 
the background of experimental and data analysis methods. 
Sections~\ref{sec:3} and \ref{SEC-DISCUSSION}
present the results and discussion of the data analysis, respectively.
The summary and outlook are described in Section 
\ref{SEC-SUMMARY}.


\section{Method \label{SEC-METHOD}}

\subsection{CTS measurements}
\label{sec:cts_measurements}

The CTS measurement was carried out in non-equilibrium plasmas generated by laser-powered capacitor coils to investigate magnetic reconnection at low-$\beta$ \cite{zhang23nphys,ji24pop}. Heating one of the capacitor plates with high-power lasers generates a strong electrostatic potential, which drives strong currents along the U-shaped coil pair. The magnetic fields between the coils reconnect, ejecting plasma outflows from the reconnection region. The CTS measurement was performed \SI{600}{\micro\meter} away from the coil center, with the scattering vector aligned nearly along the propagation direction of the electron outflow \cite{zhang23nphys}. The measurement volume was $\SI{60}{\micro\meter} \times \SI{60}{\micro\meter} \times \SI{50}{\micro\meter}$. The probe laser operated at \SI{526.5}{\nano\meter} with a scattering angle of \SI{63}{\degree}. The spectral ranges of the spectrometers were \SI{340}{\nano\meter} for the EPW feature and \SI{7}{\nano\meter} for the IAW feature \cite{katz12rsi}. In Ref.~\cite{ji24pop}, the IAW and EPW features were fitted separately; the electron temperature and density were inferred from the EPW feature and then fixed in the fit to the IAW feature to obtain the remaining plasma parameters.

Additionally, the EPW feature shows non-ideal peaks that correspond to the electron acoustic wave activities during magnetic reconnection \cite{zhang23nphys}. \rev{The accelerated electrons downstream of the reconnection region distort the electron distribution function, resulting in a non-Maxwellian distribution function that modifies the dispersion relation and Landau damping rate of electrostatic fluctuations and gives rise to additional peaks in the CTS spectrum \cite{sakai26pop}.} 

\begin{figure}[t]
    \centering
    \includegraphics[width=0.5\textwidth]{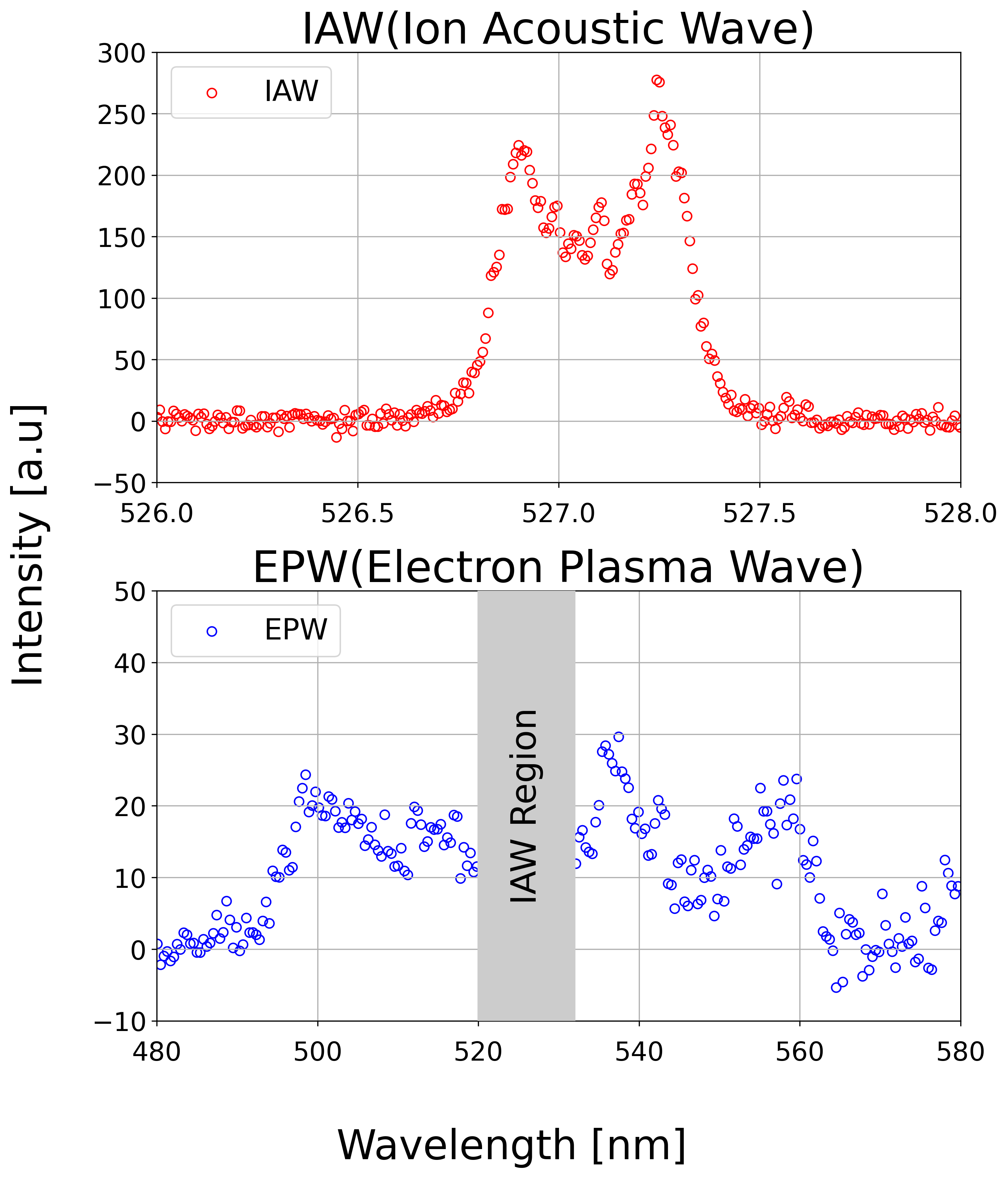}
    \caption{CTS spectra obtained at 2.9~ns after drive laser irradiation \cite{ji24pop}. The upper and lower panels show the IAW and EPW features, respectively.}
    \label{fig:IAW_and_EPW}
\end{figure}

In the present Bayesian analysis, we use spectra of both the EPW and IAW features obtained at \SI{2.9}{\nano\second} after drive laser irradiation. The background-subtracted data, corrected for the spectral response, are shown in Fig.~\ref{fig:IAW_and_EPW}. 

\subsection{Data analysis method based on Bayesian inference}


Following standard Bayesian inference
\cite{bishop2006prml,vontoussaint2011bayesian},
the posterior probability density function (PDF) $P(X|Y)$
can be computed as
\begin{eqnarray}
P(X|Y) = \frac{P(Y|X)P(X)}{P(Y)},
\label{EQ-BayesTheorem}
\end{eqnarray}
where $Y \equiv (Y_1, Y_2, \ldots, Y_m)$ is the observed data vector and
$X \equiv (X_1, X_2, \ldots, X_n)$ is the target parameter vector.
The main issue in Bayesian modeling is constructing an appropriate likelihood function $P(Y|X)$.
The prior probability density $P(X)$ is assumed to be uniform,
i.e., $P(X) \equiv \mathrm{const.}$, in a defined region of the
parameter space, as is often performed \cite{motoyama22cpc}. 
The model evidence $P(Y)$ is the normalization factor determined by $\int P(X|Y)dX=1$.
\AN{The observed data vector \(Y=(Y_1,Y_2,\ldots,Y_m)\) represents the experimentally observed CTS spectrum at the wavelength points \((\lambda_1,\lambda_2,\ldots,\lambda_m)\), with \(Y_i=I_{\rm exp}(\lambda_i)\).
The parameter vector $X$ consists of the six plasma parameters:}
the electron temperature $T_{\rm e}$, 
the electron density $n_{\rm e}$, 
the electron velocity $v_{\rm e}$,
the ion temperature $T_{\rm i}$, 
the ion velocity $v_{\rm i}$, and 
the ion charge state $Z$, 
$X = (X_1, X_2, \ldots, X_6) = (T_{\rm e}, n_{\rm e}, v_{\rm e}, T_{\rm i}, v_{\rm i}, Z)$.

\rev{
In the present study, we consider three analysis configurations:
an analysis using only the IAW feature, an analysis using only the
EPW feature, and a combined analysis using both features.
These are hereafter referred to as the IAW-only, EPW-only, and
combined analyses, respectively.
The detailed definitions of the residual functions used in these
analyses are given in Secs.~\ref{sec:3.1}, \ref{sec:3.2}, and
\ref{sec:3.3_combine}.
}

\AN{The parameter ranges used in this analysis are}
$100 \le T_{\rm e} \le 500$ eV, 
$0 \le n_{\rm e} \le 6 \times 10^{18}\,\mathrm{cm}^{-3}$, 
$-1200 \le v_{\rm e} \le -200$ km/s, 
$750 \le T_{\rm i} \le 3000$ eV, 
$-400 \le v_{\rm i} \le -250$ km/s, 
$10 \le Z \le 30$, which are set around the result obtained from the conventional analysis \cite{zhang23nphys,ji24pop}. 
\marker{These ranges define the support of the uniform prior distribution. If the posterior PDF or the optimal parameter point is concentrated near the boundary of the assumed range, this may indicate that the range is insufficient or that the data and model do not properly constrain the parameter.}
\rev{
To examine the dependence on the prior ranges, we additionally
repeated the analyses using broader ranges of the uniform priors.
The widest prior ranges examined were
$100 \le T_{\rm e} \le 900$ eV,
$0 \le n_{\rm e} \le 8 \times 10^{18}\,\mathrm{cm}^{-3}$,
$-1800 \le v_{\rm e} \le -200$ km/s,
$750 \le T_{\rm i} \le 3000$ eV,
$-400 \le v_{\rm i} \le -250$ km/s, and
$10 \le Z \le 50$.
Compared with the original prior ranges, the ranges of
$T_{\rm e}$, $n_{\rm e}$, $v_{\rm e}$, and $Z$ were substantially
broadened, whereas those of $T_{\rm i}$ and $v_{\rm i}$ were unchanged.
These calculations were performed to assess the robustness of the
inference against the assumed prior ranges.
}

The ion density $n_{\rm i}$ is determined by charge neutrality, $n_{\rm i} = n_{\rm e}/Z$. 
The model CTS spectral intensity $I_{\rm cal}(\lambda;X)$ at each wavelength point is calculated using the dynamic structure factor \cite{froula11}. In the present analysis, the electron and ion velocity distribution functions are assumed to be Maxwellian.
Because the absolute scales of the calculated spectral intensity $I_{\rm cal}(\lambda;X)$ and the observed intensity $I_{\rm exp}(\lambda)$ are different, the two spectra are compared after appropriate normalization. 
We use a standard form of the normalization condition as 
\begin{eqnarray}
\sum_{i=1}^{m} I_{\rm cal}(\lambda_i;X)  =
\sum_{i=1}^{m} I_{\rm exp}(\lambda_i)  = 1,
\label{EQ-STD-NORM-COND}
\end{eqnarray}
\AN{which is referred to as the ``uniform'' normalization condition in this paper.}
\rev{
The wavelength range used for this normalization depends on the analysis
configuration, as described in Secs.~\ref{sec:3.1}--\ref{sec:3.3_combine}.
}

The likelihood function for the observed spectrum at a given wavelength $(\lambda=\lambda_i)$ is modeled as a Gaussian function, 
\begin{eqnarray}
P(I_{\rm exp}(\lambda_i)|X) \propto 
\exp \left[
- \beta
\left(
I_{\rm cal}(\lambda_i;X) - I_{\rm exp}(\lambda_i)
\right)^2
\right],
\label{EQ:gaussian}
\end{eqnarray}

\rev{
In the present analysis, the residuals between the observed and
calculated intensities are modeled as independent Gaussian random variables
with a common effective variance after spectral normalization. With the
definition in Eq.~\eqref{EQ:gaussian}, this effective residual variance is
$(2\beta)^{-1}$.
Accordingly, $\beta$ is interpreted as an effective inverse
residual-variance parameter rather than as a direct estimate of the
detector-noise variance. The present likelihood is therefore regarded as
an effective residual model; wavelength-dependent or correlated measurement
uncertainties are not explicitly included.
}

Then, the likelihood function for the whole observed spectrum
is set to be  
\begin{eqnarray}
P(Y|X) & \propto &  
\prod_{i=1}^{m} P(I_{\rm exp}(\lambda_i)|X) \nonumber \\
& \propto & 
\exp \left[
- \beta \sum_{i=1}^{m}
\left(
I_{\rm cal}(\lambda_i;X) - I_{\rm exp}(\lambda_i)
\right)^2
\right] \nonumber \\
&=& \exp \left( - \beta F(X;Y) \right),
\end{eqnarray}
with the residual difference function $F(X;Y)$ defined by 
\begin{eqnarray}
F(X;Y) \equiv
\sum_{i=1}^{m}
\left(
I_{\rm cal}(\lambda_i;X) - I_{\rm exp}(\lambda_i)
\right)^2.
\end{eqnarray}

\rev{
Because the likelihood depends on $\beta$, the model evidence is
denoted by $P(Y;\beta)$. The hyperparameter $\beta$ is selected from
the maximum of the model evidence, whose explicit definition is given
in Sec.~\ref{sec:3}. The model evidence is evaluated using the fully
normalized Gaussian likelihood, including its $\beta$-dependent
normalization factor.
}

\marker{We use} 
the population annealing Monte Carlo (PAMC) method
\cite{hukushima03aipp} implemented in the data-analysis framework 
ODAT-SE \cite{motoyama22cpc}.
The PAMC method is 
a parallelized Monte Carlo method suitable for 
modern massively parallel supercomputers.
See Ref.~\cite{motoyama22cpc} for further details.
\marker{We choose the PAMC method because it can sample the full high-dimensional posterior distribution, including parameter correlations and possible multiple solution candidates, while being suitable for massively parallel computation.}

The PAMC analysis gives $P(X|Y)$
as a histogram in the six-dimensional parameter space.
A sampling point in the histogram is referred to as a replica, and 
the number of replicas $N_{\rm replica}$ 
determines the reliability of $P(X|Y)$.
The sampling was performed using a discretized hyperparameter schedule with
$N_{\beta}=61$ points distributed logarithmically in the range
$1 \le \beta \le 10^{7}$.
At each value of $\beta$, $N_{\rm MC}=30$ Monte Carlo updates were carried out.
The number of replicas was set to $N_{\rm replica}=76,800$, and the computational time was about 3 hours using 384 nodes of Fugaku, the Japanese flagship supercomputer.
\rev{
As a stability and reproducibility check, the combined analysis
was repeated using different random seeds. The resulting
model-evidence curves and their evidence-maximizing regions were
consistent among the independent runs, indicating that the PAMC
sampling is sufficiently stable with respect to the random seed.
}

\AN{The marginal posterior PDF for parameter $X_i$ is calculated as
\begin{eqnarray}
P(X_i|Y) \equiv \int \, \delta (X'_i - X_i) \, P(X'|Y) \, dX',
\end{eqnarray}
where $i=1,2,\ldots,6$ and
$X' \equiv (X_1', X_2', \ldots, X_6')$.}
\rev{
As a sensitivity check of the prior ranges, we compared the
marginal posterior PDFs obtained with the original and broadened
uniform priors described above.
Although the detailed shapes of the marginal posterior PDFs were
affected to some extent by the prior boundaries, no clearly isolated
peak in $Z$ appeared in either the IAW-only or EPW-only analysis
over the prior ranges examined, whereas $Z$ remained clearly
constrained in the combined analysis.
Thus, the main conclusion that the complementary information from
the IAW and EPW features is essential for constraining $Z$ is robust
over the prior ranges examined here.
}

\section{Results}
\label{sec:3}

The CTS spectra were obtained using two spectrometers with different wavelength ranges and resolutions. We refer to the spectra obtained with the high-resolution and broadband spectrometers as the IAW and EPW features, respectively, because these spectra correspond to the dispersion characteristics of IAWs and EPWs. We analyzed the data within the wavelength ranges $526\le\lambda\le 528$~nm for the IAW feature, and $480\le\lambda\le 580$~nm for the EPW feature, as shown in Fig.~\ref{fig:IAW_and_EPW}. To avoid the influence of the strong IAW signal on the analysis of the EPW feature, the data within $520\le \lambda \le 530$~nm were excluded from the EPW analysis.

\AN{Using the CTS spectral model \cite{froula11}, we performed three types of analyses: (i) an IAW-only analysis, (ii) an EPW-only analysis, and (iii) a combined analysis in which the IAW and EPW spectra were normalized separately and fitted jointly. The results are shown in Figs.~\ref{fig:IAW}, \ref{fig:EPW}, and \ref{fig:combine}, respectively.
\rev{
The hyperparameter $\beta$ was determined on the basis of the model evidence~\cite{friel12sn}.
For a given value of $\beta$, the model evidence is defined as the marginal likelihood obtained by integrating the likelihood over the parameter space,
\begin{equation}
P(Y;\beta)
=
\int P(Y|X;\beta) P(X)\,dX .
\label{eq:model_evidence_beta}
\end{equation}

\begin{figure}[t]
    \centering
    \includegraphics[width=0.5\textwidth]{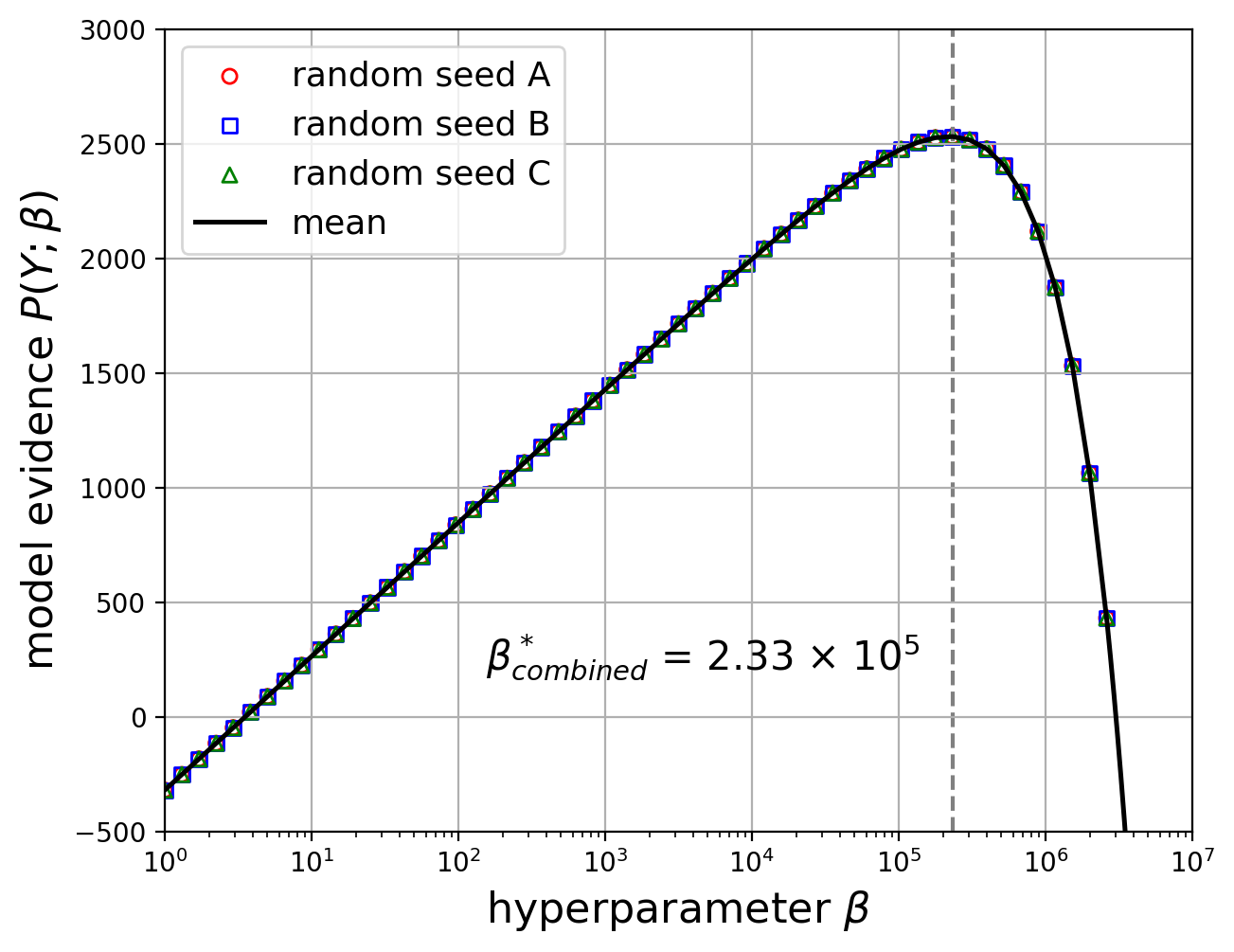}
    \caption{\rev{
    Model evidence $P(Y;\beta)$ for the combined analysis obtained from three independent PAMC runs with different random seeds.
    The three runs show consistent $\beta$ dependence and the same evidence-maximizing region around
    $\beta_{\rm combined}^{*}=2.33\times10^{5}$,
    demonstrating the reproducibility of the PAMC sampling.
    }}
    \label{fig:evidence_seed}
\end{figure}

The model evidence quantifies the overall consistency between the observed data and the model over the parameter space for a given $\beta$.
In the combined analysis, the model evidence takes its maximum at
\begin{equation}
\beta_{\rm combined}^{\ast}
=
\mathop{\mathrm{arg\,max}}\limits_{\beta} P(Y;\beta)
=
2.33\times10^{5}.
\label{eq:beta_opt}
\end{equation}
For a direct comparison of the posterior structures using the same value of
$\beta$, we use $\beta_{\rm combined}^{\ast}$ for the
posterior PDFs shown in Figs.~\ref{fig:IAW}, \ref{fig:EPW}, and
\ref{fig:combine}.
The evidence-maximizing values obtained independently from the IAW-only
and EPW-only analyses differ because the three analyses use different
data sets and residual functions.
}

\marker{For each analysis, the optimal parameter point is defined as}
\(X^{\ast} \equiv \arg \min_X F(X;Y)\), \marker{where} \(F(X;Y)\) \marker{denotes the residual difference function used in the corresponding analysis. The model spectra shown below are evaluated at} \(X^{\ast}\).
In Figs.~\ref{fig:IAW}, \ref{fig:EPW}, and \ref{fig:combine}, panel (a) shows the one-dimensional (1D) marginal posterior PDFs of the six plasma parameters. The red distributions represent the 1D marginal posterior PDFs, and the black dotted vertical lines indicate the optimal parameter values. Panels (b) and (c) compare the experimental and model spectra. In panel (b), the red points show the IAW experimental data, and the black dotted line shows the model spectrum evaluated at the optimal parameter point. In panel (c), the blue points show the EPW experimental data, and the black dotted line shows the model spectrum evaluated at the same optimal parameter point.}

It should be noted that the optimal parameter point
\(X^*\) in the full six-dimensional
parameter space does not necessarily coincide with the peak
positions of the \AN{1D marginal posterior PDFs}.
\AN{The peak positions of the 1D marginal posterior PDFs are obtained by
marginalizing the full posterior PDF onto each parameter axis. They therefore
include the effects of the volume and correlation structure of the
high-dimensional posterior PDF. 
\rev{
In other words, the optimal solution identifies the single most favorable point in the full parameter space, whereas the peak of a 1D marginal posterior reflects where the total posterior probability is most concentrated after integrating over the other parameters.
}
Consequently, the model spectrum evaluated at
\(X^*\) and the peak positions of the 1D marginal posterior PDFs
characterize different aspects of the inference result; they are complementary
but do not necessarily coincide.}




\subsection{Analysis with the ion acoustic wave (IAW) data set}
\label{sec:3.1}



\begin{figure}[t]

    \centering

    \includegraphics[width=0.49\textwidth]{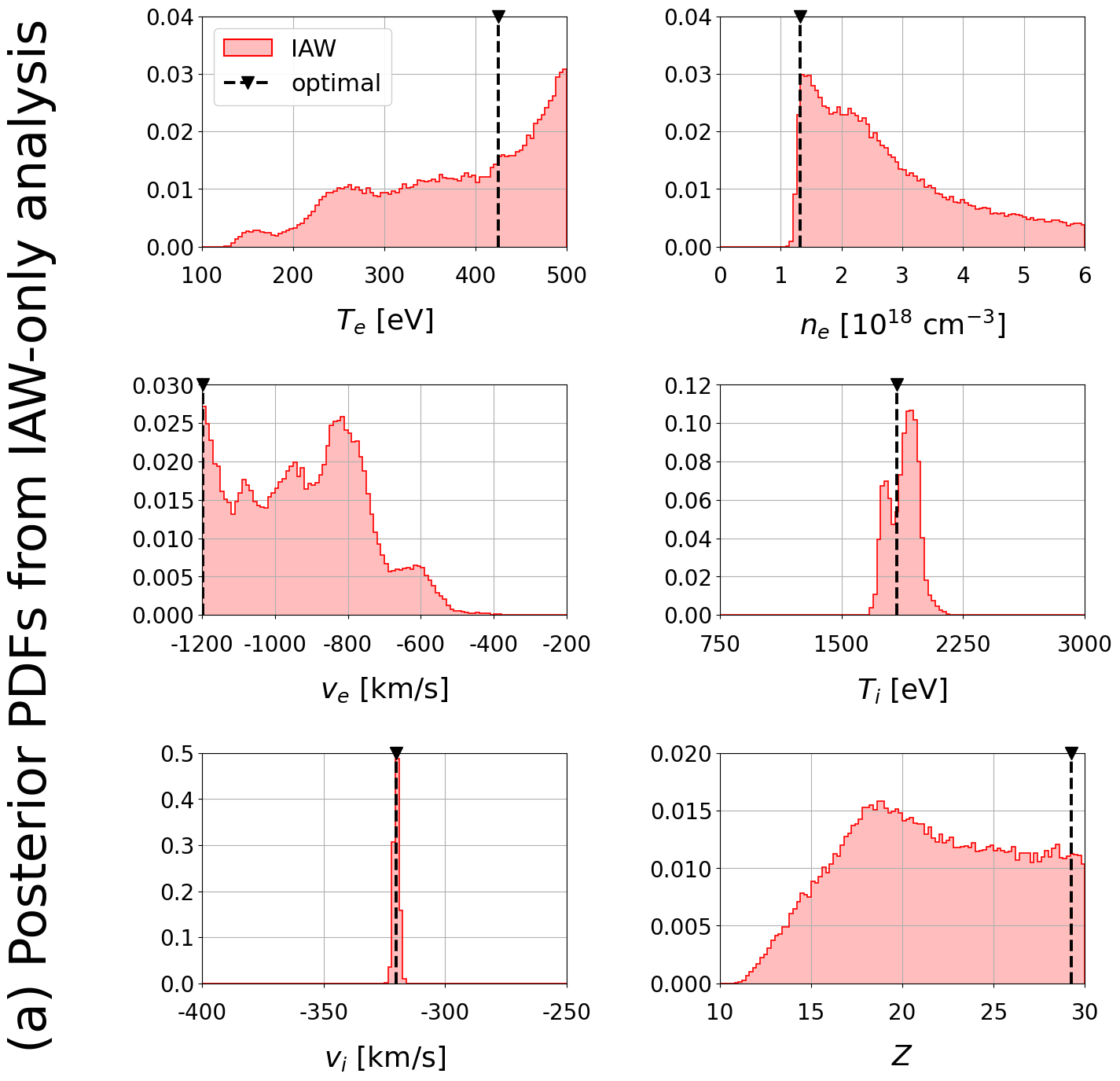}

    \vspace{2mm}

    \includegraphics[width=0.35\textwidth]{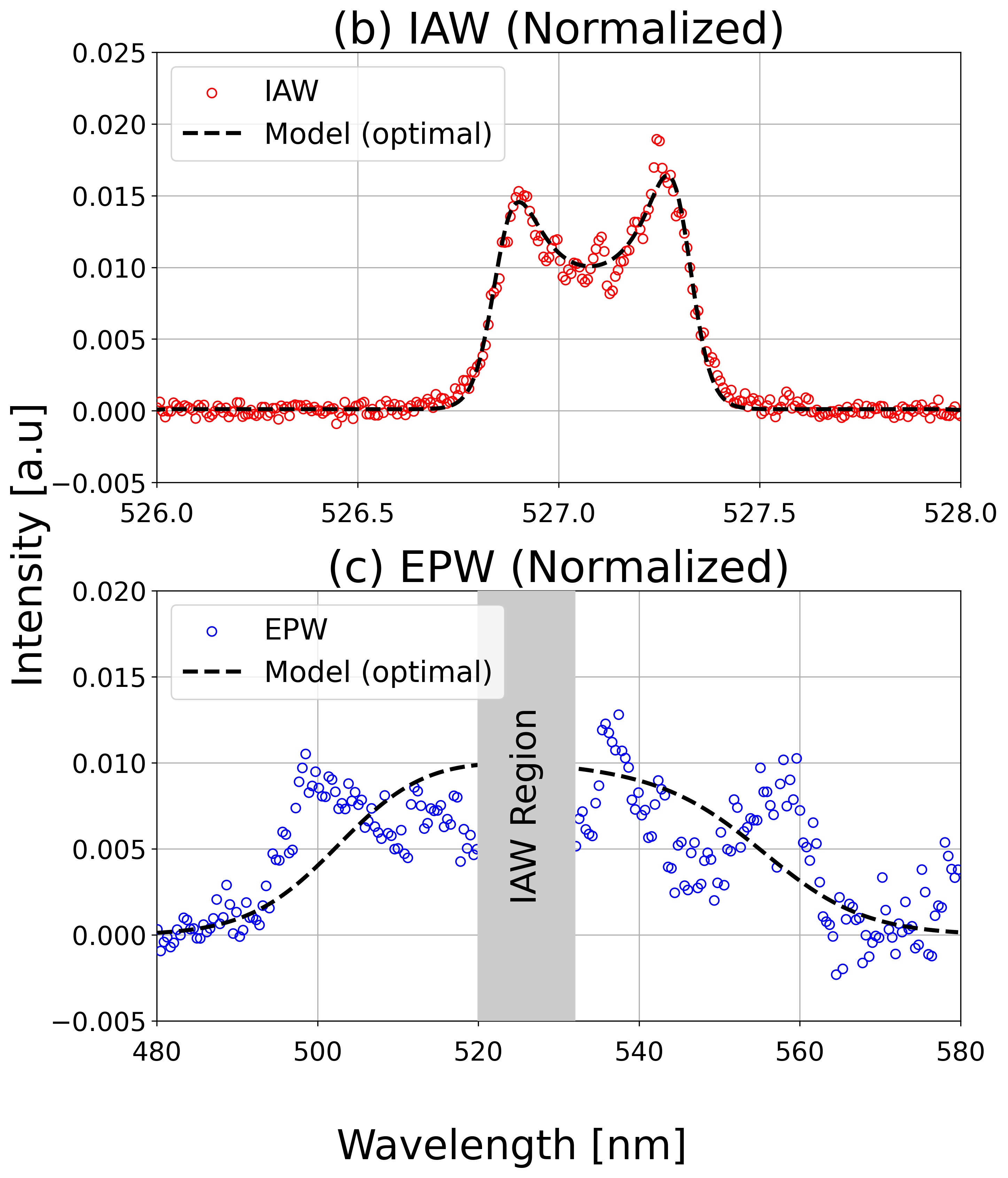}

    \caption{\AN{Result of the IAW-only analysis at $\beta = 2.33 \times 10^5$: (a) 1D marginal posterior PDFs, (b) IAW spectrum, and (c) EPW spectrum.}}
    \label{fig:IAW}
\end{figure}

\AN{Figure~\ref{fig:IAW} shows the result of the IAW-only analysis. Here, the residual difference function 
in the likelihood function is defined as}
\begin{eqnarray}
F(X;Y) &:=& F_{\rm IAW}(X;Y) \nonumber \\
&\equiv&
\sum_{\lambda_i \in {\rm IAW}}
\left[
I_{\rm cal}(\lambda_i;X) - I_{\rm exp}(\lambda_i)
\right]^2
\end{eqnarray}
\AN{under the uniform normalization condition applied only to the IAW feature,} 
\begin{eqnarray}
\sum_{\lambda_i \in {\rm IAW}} I_{\rm cal}(\lambda_i;X)
=
\sum_{\lambda_i \in {\rm IAW}} I_{\rm exp}(\lambda_i)
= 1.
\label{EQ-IAW-NORM-COND}
\end{eqnarray}
\AN{Here, the summation is carried out only over the IAW feature} (526 nm $\le \lambda \le $ 528 nm).
\AN{The model spectrum evaluated at the optimal parameter point reproduces the IAW data, whereas discrepancies remain in the EPW feature.}
In the \AN{1D marginal posterior PDFs}, the distributions of $T_{\rm i}$ and $v_{\rm i}$ show relatively narrow peaks.
By contrast, the distributions of $T_{\rm e}$, $n_{\rm e}$, $v_{\rm e}$, and $Z$ are broader.
In particular, the distribution of $Z$ does not show a clearly isolated peak in the IAW-only analysis.




\subsection{Analysis with the electron plasma wave (EPW) data set}
\label{sec:3.2}

\begin{figure}[t]  
    \centering

    \includegraphics[width=0.49\textwidth]{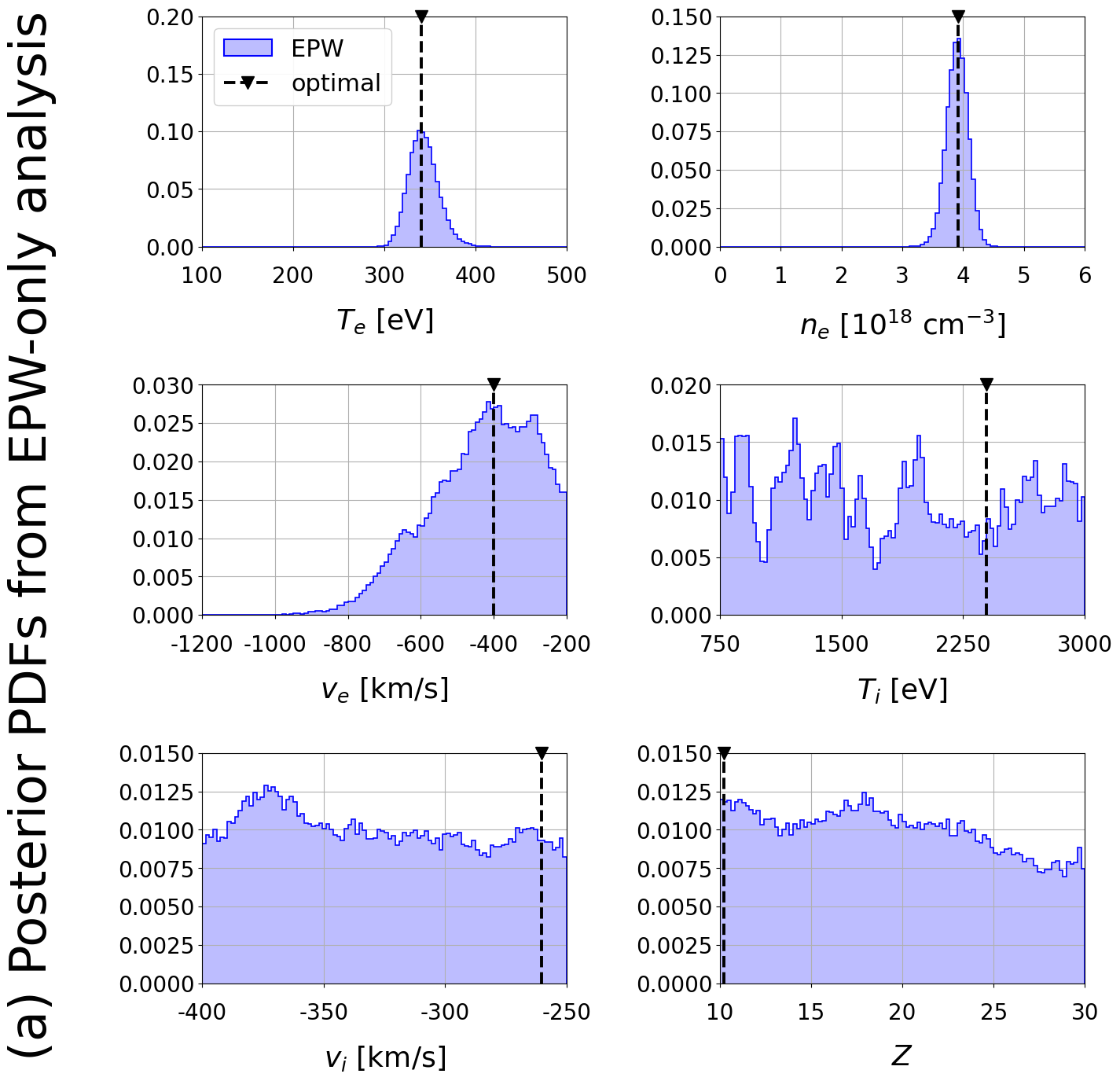}

    \vspace{2mm}
    \includegraphics[width=0.35\textwidth]{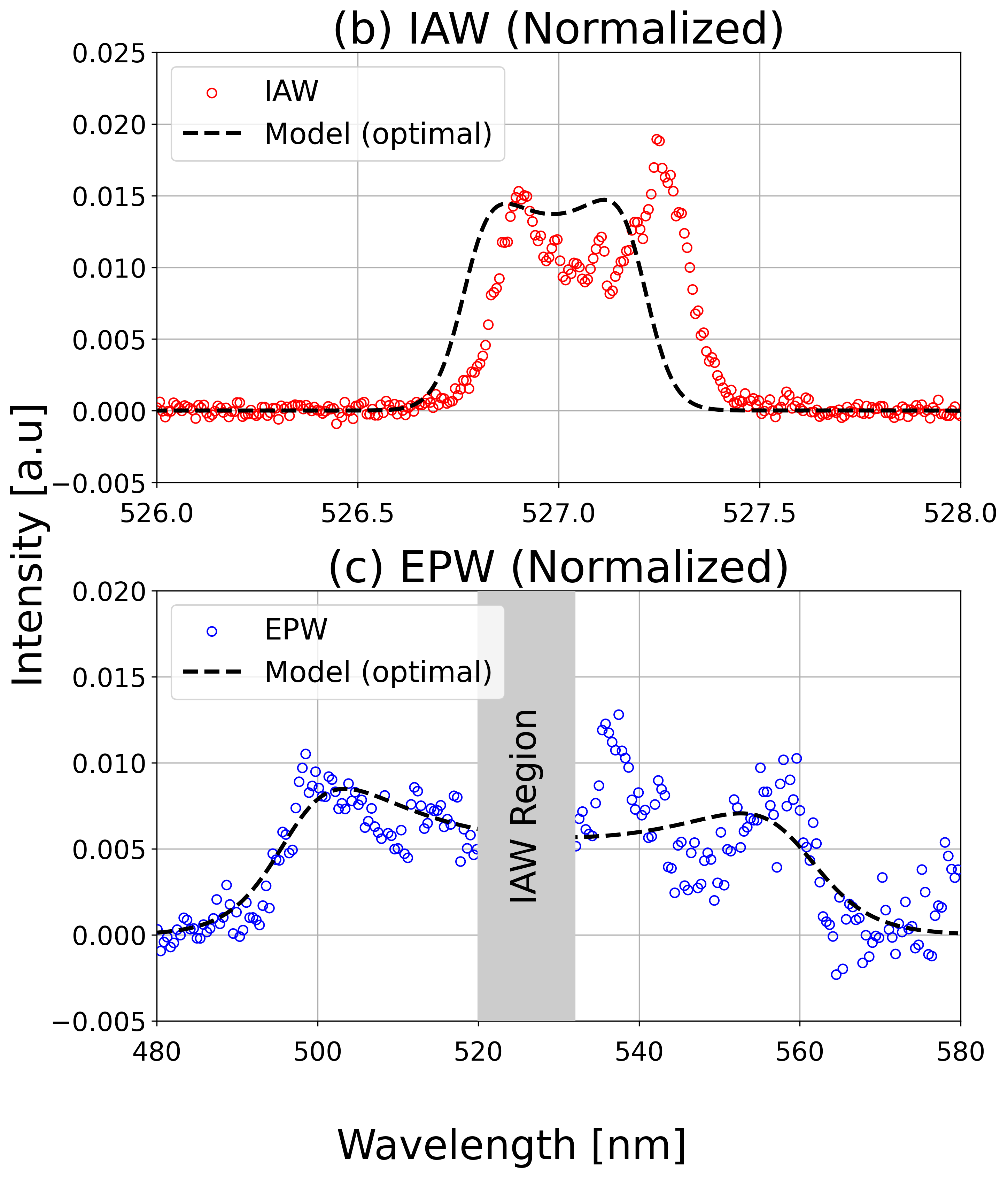}


    \caption{\AN{Result of the EPW-only analysis at $\beta = 2.33 \times 10^5$: (a) 1D marginal posterior PDFs, (b) IAW spectrum, and (c) EPW spectrum.}}
    \label{fig:EPW}
\end{figure}



\AN{Figure~\ref{fig:EPW} shows the result of the EPW-only analysis. The EPW-only analysis was performed under the uniform normalization condition applied to the EPW feature,
\begin{eqnarray}
\sum_{\lambda_i \in {\rm EPW}} I_{\rm cal}(\lambda_i;X)
=
\sum_{\lambda_i \in {\rm EPW}} I_{\rm exp}(\lambda_i)
= 1.
\label{EQ-EPW-NORM-COND}
\end{eqnarray}
The residual difference function for the PAMC analysis is defined as
\begin{eqnarray}
F(X;Y) &:=& F_{\rm EPW}(X;Y) \nonumber \\
&\equiv&
\sum_{\lambda_i \in {\rm EPW}}
\left[
I_{\rm cal}(\lambda_i;X) - I_{\rm exp}(\lambda_i)
\right]^2,
\end{eqnarray}
without the contribution from the IAW data.}

\AN{The model spectrum evaluated at the optimal parameter point reproduces the main EPW feature, whereas discrepancies remain in the IAW feature.}
In the \AN{1D marginal posterior PDFs}, the distributions of $T_{\rm e}$ and $n_{\rm e}$ show narrow peaks.
The distribution of $v_{\rm e}$ has a broader structure with multiple local maxima.
The distributions of $T_{\rm i}$, $v_{\rm i}$, and $Z$ are broader than those of $T_{\rm e}$ and $n_{\rm e}$.
Additional structures are observed in the experimental EPW data around $\sim 515$~nm and $\sim 535$~nm, which are not fully reproduced by the present Maxwellian model.

\subsection{Combined analysis with IAW and EPW data sets}
\label{sec:3.3_combine}

\begin{figure}[t]
    \centering

    \includegraphics[width=0.49\textwidth]{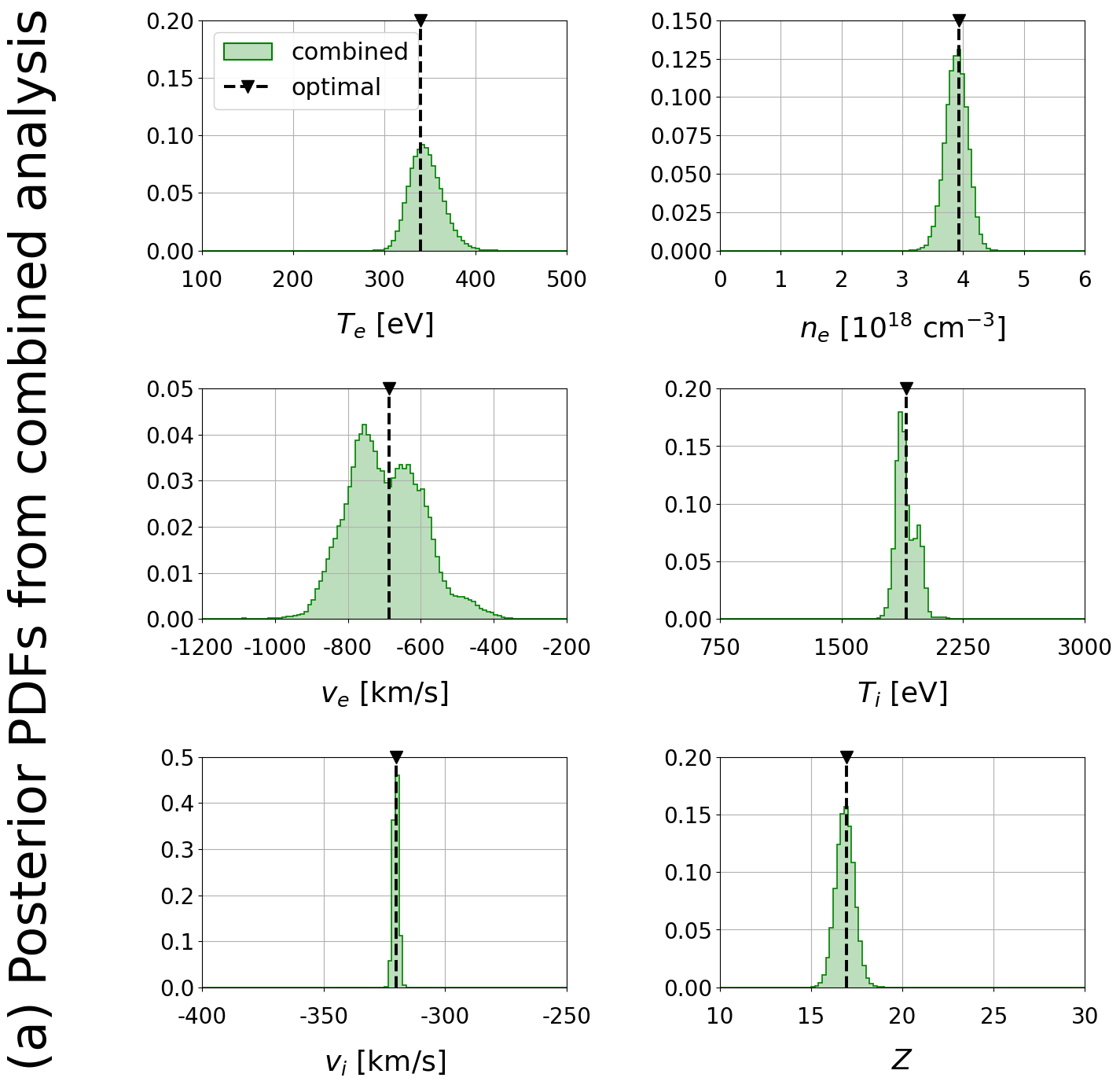}

    \vspace{2mm}

    \includegraphics[width=0.35\textwidth]{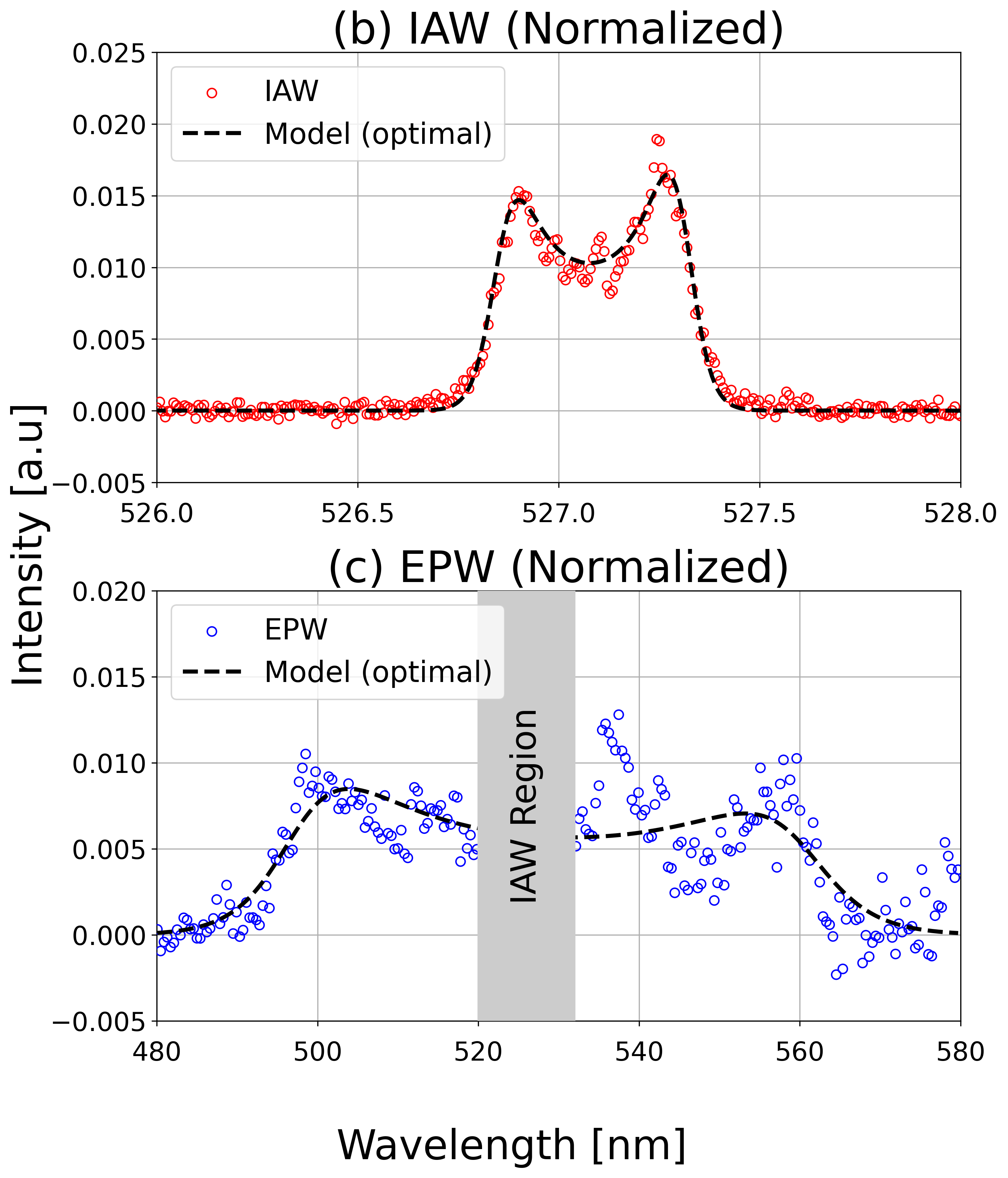}

    \caption{\AN{Result of the combined analysis at $\beta = 2.33 \times 10^5$: (a) 1D marginal posterior PDFs, (b) IAW spectrum, and (c) EPW spectrum.}}
    \label{fig:combine}
\end{figure}


In the combined analysis, the IAW and EPW spectra were independently normalized and fitted jointly.
The normalization and construction of the combined residual are described by

%
\AN{
\begin{equation}
\begin{aligned}
&F_{\mathrm{combined}}(X;Y) \\
&\quad =
\frac{1}{2}
\left[
F_{\mathrm{IAW}}(X;Y)
+
F_{\mathrm{EPW}}(X;Y)
\right].
\end{aligned}
\label{equ:combined}
\end{equation}
}


Figure~\ref{fig:combine} shows the result of the combined analysis.
\AN{The model spectra evaluated at the optimal parameter point reproduce both the IAW and EPW features.}
\AN{In the 1D marginal posterior PDFs, the combined analysis provides constraints on all six parameters, although the degree of constraint depends on the parameter.}
In particular, \AN{the posterior PDF} of the ion charge state $Z$ exhibits a peak around $Z\sim 17$, whereas such a clear peak is not obtained in either the IAW-only or EPW-only analysis.
This value is close to the averaged ion charge state obtained from the collisional radiative model \cite{chung05hedp}.

It should be noted that the separate normalization conditions in Eqs.~
\eqref{EQ-IAW-NORM-COND}
 and 
\eqref{EQ-EPW-NORM-COND}
 are crucial for capturing the two-component nature of plasmas, consisting of ions and electrons. A numerical demonstration of their importance is given in Appendix~\ref{app:all_region}.
\rev{
In Eq.~\eqref{equ:combined}, the separately normalized IAW and EPW
residuals are assigned equal weights. The sensitivity to this weighting
choice is examined using data-point weighting in
Appendix~\ref{app:data_point_weighting}.
}



\section{Discussion \label{SEC-DISCUSSION}}

\subsection{Data-driven identification of sensitivity}

\AN{The parameter sensitivities are visualized by the sharpness of the 1D marginal posterior PDFs.} As shown in Fig.~\ref{fig:IAW}, the IAW feature contains information on all parameters, as all posterior \AN{PDF}s show non-uniformity. In contrast, Fig.~\ref{fig:EPW} shows that the EPW feature does not constrain the ion parameters ($T_{\rm i}$, $v_{\rm i}$, and $Z$) due to the flat posterior \AN{PDF}s. 

As explained in many plasma physics textbooks (e.g., Ref.~\cite{chen16}), ions act as a nearly stationary background that supports charge separation in high-frequency EPWs; consequently, ion dynamics has a negligible effect on the characteristics of EPWs. In contrast, electrons respond rapidly to the motion of ions to maintain quasi-neutrality in low-frequency IAWs, and electron dynamics plays an important role in the pressure balance. Therefore, the sensitivities of EPW and IAW features obtained from Bayesian inference are physically reasonable, even though Bayesian inference does not use any information other than that contained in the dynamic structure factor. This suggests that Bayesian inference can reveal the physical implications encoded in the equations, which is particularly important for analyzing the physics behind complex models of physical phenomena. In particular, this capability facilitates the interpretation of the characteristic shapes of the velocity distribution functions in the CTS analysis assuming non-Maxwellian velocity distribution functions \cite{sakai20pop,sakai23pop}.

\subsection{Impact of combined analysis}
\label{sec:4.2}
\begin{figure}[t]
    \centering
    \includegraphics[width=\hsize]{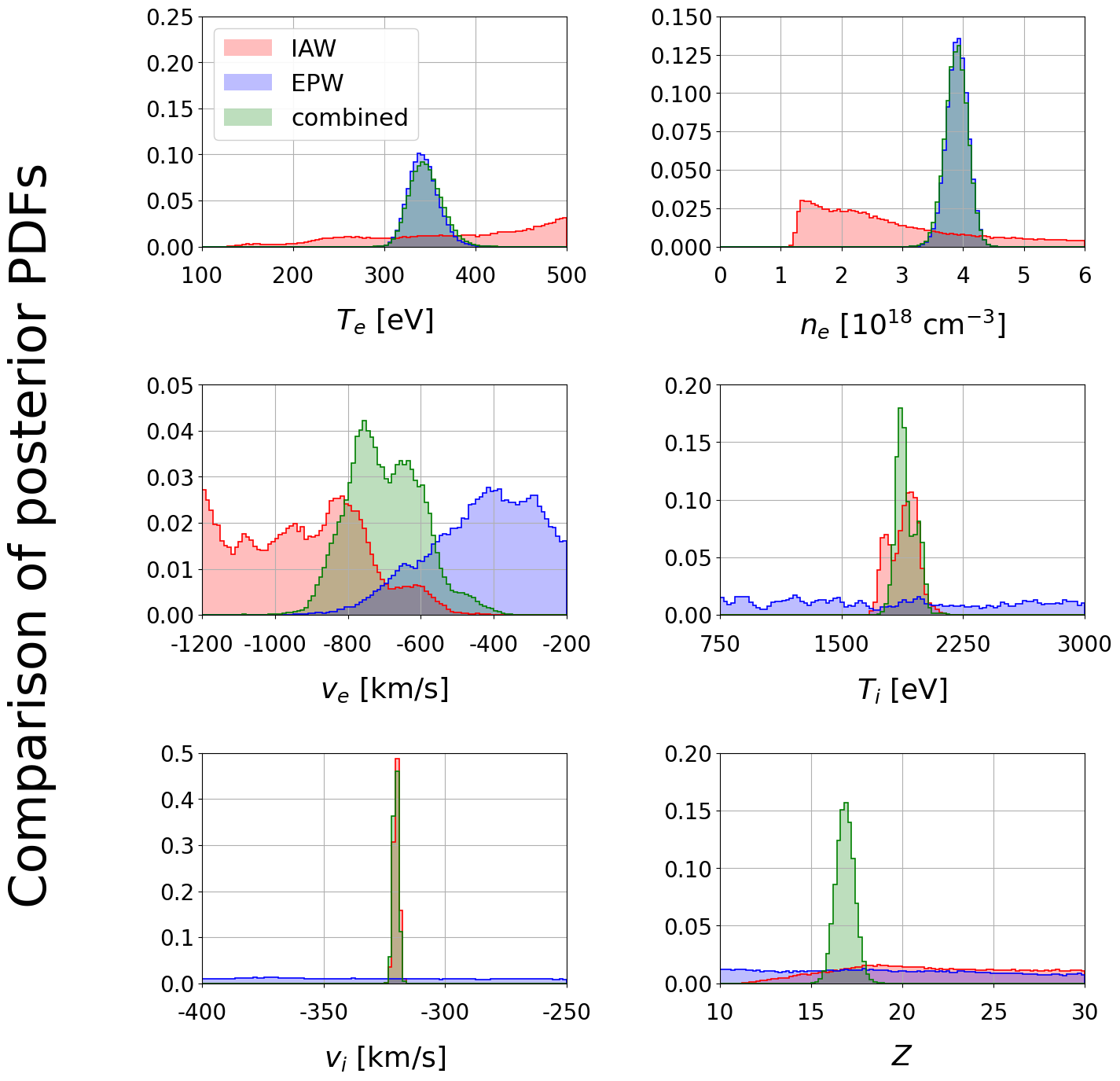}
    \caption{\AN{Comparison of 1D marginal posterior PDFs obtained from the IAW-only, EPW-only, and combined analyses.}}
    \label{fig:1dpdf}
\end{figure}
%

Figure~\ref{fig:1dpdf} compares the \AN{1D marginal posterior PDFs} obtained from the IAW-only, EPW-only, and combined analyses.
The estimated electron parameters, particularly the electron temperature, electron density, and electron velocity, exhibit noticeable discrepancies between the EPW-only and IAW-only analyses.
These discrepancies reflect the different parameter sensitivities and noise characteristics of the two spectral features.
In contrast, the combined analysis yields narrower posterior PDFs for all six parameters.


\begin{figure}[t]
    \centering
    \includegraphics[width=\hsize]{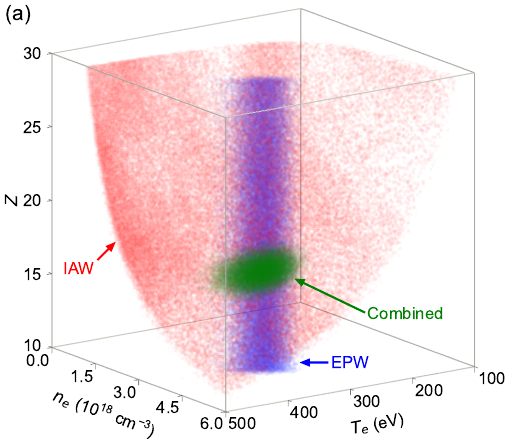}
    \includegraphics[width=\hsize]{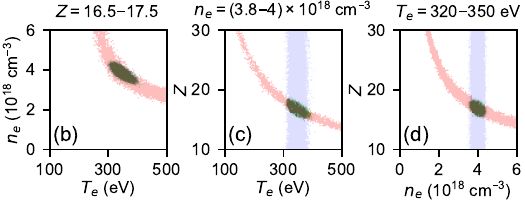}
    \caption{\AN{(a) Comparison of 3D marginal posterior PDFs of $T_{\rm e}$, $n_{\rm e}$, and $Z$ obtained from the IAW-only, EPW-only, and combined analyses.} (b)--(d) 2D slices of the 3D marginal posterior PDFs around the optimal parameter point.}
    \label{fig:3dpdf}
\end{figure}

\rev{
Figures~\ref{fig:combine}--\ref{fig:3dpdf} provide complementary views of how the ion charge state $Z$ is constrained by the combined analysis. Figure~\ref{fig:combine} shows that the combined analysis produces a clear peak in the 1D marginal posterior PDF of $Z$ around $Z\sim17$, together with constrained posterior PDFs for the other plasma parameters. Figure~\ref{fig:1dpdf} compares the IAW-only, EPW-only, and combined analyses and demonstrates that such a clear constraint on $Z$ appears only in the combined analysis. Figure~\ref{fig:3dpdf} reveals the origin of this constraint in the high-dimensional parameter space: the IAW-only posterior forms a correlated structure among $T_{\rm e}$, $n_{\rm e}$, and $Z$, whereas the EPW-only posterior strongly constrains $T_{\rm e}$ and $n_{\rm e}$ but remains weakly constrained along the $Z$ direction. Their combination localizes the posterior near the intersection of these complementary constraint structures, thereby breaking the degeneracy in $Z$ and producing the clear peak observed in Figs.~\ref{fig:combine} and \ref{fig:1dpdf}.}

Figure~\ref{fig:3dpdf}(a) compares the \AN{three-dimensional (3D)} posterior \AN{PDF}s of electron temperature, electron density, and ion charge state in the IAW-only (red), EPW-only (blue), and combined (green) analyses. The two-dimensional (2D) slices of Fig.~\ref{fig:3dpdf}(a) are shown in Figs.~\ref{fig:3dpdf}(b)--\ref{fig:3dpdf}(d). The parameter ranges to obtain the slices in Figs.~\ref{fig:3dpdf}(b)--\ref{fig:3dpdf}(d) are $16.5 \le Z \le 17.5$, $\SI{3.8e18}{cm^{-3}} \le n_e \le \SI{4e18}{cm^{-3}}$, and $\SI{320}{eV} \le T_e \le \SI{350}{eV}$, respectively.
The IAW-only result exhibits a curved surface in the 3D parameter space, indicating a correlation among these parameters.
The EPW-only result constrains the electron temperature and density, producing \AN{a distribution elongated along the \(Z\) direction.}
The \AN{3D} posterior \AN{PDF} obtained from the combined analysis is localized near the intersection of \AN{this elongated distribution and the curved surface.}
Therefore, combining the high-dimensional correlations in the IAW data with the constraints from the EPW data leads to the formation of a clear peak in the posterior \AN{PDF} of the ion charge state.
The 3D correlation observed in the IAW-only analysis can be explained as follows.
The phase speed of the IAWs is given by \AN{$c_s = [(Z T_{\rm e} + 3 T_{\rm i})/m_{\rm i}]^{1/2} \approx (Z T_{\rm e}/m_{\rm i})^{1/2}$}, where $m_{\rm i}$ is the ion mass.
Taking into account the condition of parametric resonance of the IAW feature, the separation between the two IAW peaks is proportional to the phase speed of the IAWs and is therefore a nonlinear function of both $T_{\rm e}$ and $Z$ \cite{sakai23pop}.
More generally, under the Salpeter approximation, the shape of the IAW spectrum is determined by $[\alpha^2/(\alpha^2 +1)] (Z T_{\rm e}/T_{\rm i})$, where $\alpha = (k \lambda_D)^{-1}$, $k$ is the scattering wavevector, and \AN{$\lambda_D \propto (T_{\rm e}/n_{\rm e})^{1/2}$} is the Debye length \cite{froula11}.
This dependence is consistent with the observed curved surface in the 3D posterior \AN{PDF} of the IAW-only analysis.
The density dependence can be understood in terms of the damping of IAWs whose wavelengths are shorter than the Debye length.
Note that the $n_{\rm e}$ dependence vanishes in the limit $\alpha \gg 1$, and the 3D correlation is reduced to a 2D correlation without $n_{\rm e}$.

A correlation among $T_{\rm e}$, $n_{\rm e}$, and $Z$ is expected in the IAW feature, whereas the EPW feature constrains $T_{\rm e}$ and $n_{\rm e}$ independently.
If the most plausible $T_{\rm e}$ and $n_{\rm e}$ inferred from the EPW feature are different from those inferred from the IAW feature, these parameters are mainly determined by the EPW feature because the widths of the posterior \AN{PDF}s in $T_{\rm e}$ and $n_{\rm e}$ from the EPW feature are narrower than those from the IAW feature.
By contrast, $Z$ is determined by the IAW feature because the EPW feature does not have sensitivity to $Z$.
To incorporate the high-dimensional \AN{correlation} that is hidden in the \AN{1D marginal posterior PDFs}, it is essential to combine the EPW and IAW analyses within the Monte Carlo sampling.
\rev{
In other words, the IAW feature primarily constrains combinations of $Z$ with the electron parameters, whereas the EPW feature independently constrains $T_{\rm e}$ and $n_{\rm e}$. Combining these complementary constraints therefore removes the degeneracy and restricts the possible range of $Z$. This degeneracy breaking is directly visualized in Fig.~\ref{fig:3dpdf}, where the combined posterior is localized near the intersection of the IAW correlation structure and the EPW constraint.}

\rev{For a quantitative comparison, at approximately 2.9 ns, the conventional analysis in Ref.~\cite{ji24pop} gives $T_{\rm e}\sim340$ eV, $n_{\rm e}\sim3.8\times10^{18}$ cm$^{-3}$, and $Z\sim18$--$19$, as estimated from Fig.~4 of Ref.~\cite{ji24pop}. In the present combined Bayesian analysis, the corresponding 1D marginal posterior PDFs have peaks around $T_{\rm e}\sim340$ eV, $n_{\rm e}\sim4.0\times10^{18}$ cm$^{-3}$, and $Z\sim17$, showing quantitative consistency with the conventional analysis. While the conventional analysis provides best-fit parameter values and their uncertainties, the present Bayesian analysis additionally provides the full posterior structure, including parameter correlations and weakly constrained directions. In particular, the high-dimensional posterior reveals how the complementary constraints from the IAW and EPW features localize the posterior distribution of $Z$, providing a statistical interpretation of its determination that is not accessible from the optimal parameter values alone.
}

The plasma parameters inferred from the combined analysis are consistent with those obtained from the conventional and automatic-differentiation-based analyses \cite{zhang23nphys,ji24pop,milder24mlst}, which include physical and instrumental effects such as gradients and finite aperture corrections. As discussed in Ref.~\cite{ji24pop}, these effects are negligible in the present data set. However, all the physical and instrumental effects that could potentially modify the model spectrum should be included to ensure the validity of the analysis for any data. The combined Bayesian modeling naturally yields parameters that are consistent between the IAW and EPW features, which is sometimes an issue in the analysis \cite{follett16rsi}. 

Moreover, the Bayesian modeling provides interpretable results through high-dimensional marginal posterior PDFs, as highlighted in Fig.~\ref{fig:3dpdf}.
\marker{The PAMC-based Bayesian approach and conventional or automatic-differentiation-based approaches have different strengths. Conventional and automatic-differentiation-based approaches are powerful when the main goal is to rapidly obtain optimal parameter values for many time-resolved spectra} \cite{milder24mlst}. \marker{In contrast, the PAMC-based Bayesian approach is useful when the full posterior structure is important, including parameter uncertainties, correlations, possible multiple solution candidates, and weakly constrained directions in parameter space }\cite{bishop2006prml,vontoussaint2011bayesian,motoyama22cpc}. \marker{In addition, because the PAMC-based approach can evaluate the model evidence} \cite{friel12sn}, \marker{it can be extended to Bayesian model selection among candidate spectral models, as discussed in non-Maxwellian CTS inference} \cite{sakai26pop}. \marker{Therefore, the analysis method should be selected according to the purpose of the study.}



\section{Summary and Outlook \label{SEC-SUMMARY}}
\AN{We applied Bayesian inference based on the population annealing Monte Carlo
(PAMC) method to the analysis of collective Thomson scattering (CTS) spectra.
The posterior probability density functions (PDFs) of six plasma parameters,
$T_{\rm e}$, $n_{\rm e}$, $v_{\rm e}$, $T_{\rm i}$, $v_{\rm i}$, and $Z$,
were evaluated for the ion acoustic wave (IAW)-only, electron plasma wave
(EPW)-only, and combined analyses.}
\AN{The IAW-only and EPW-only analyses provided complementary constraints on the
plasma parameters.
The IAW-only analysis gave narrow marginal posterior PDFs for $T_{\rm i}$
and $v_{\rm i}$, whereas the EPW-only analysis gave narrow marginal
posterior PDFs for $T_{\rm e}$ and $n_{\rm e}$.
The combined analysis provided constraints on all six parameters, although the degree of constraint depends on the parameter.
In particular, the ion charge state $Z$ was clearly constrained only in the
combined analysis.
The three-dimensional (3D) posterior PDF of $T_{\rm e}$, $n_{\rm e}$, and $Z$
showed that this determination originates from the localization of the
posterior PDF formed by combining the correlated constraint from the IAW feature
with the electron-parameter constraint from the EPW feature.}
The present combined analysis method demonstrates the importance of two-component Bayesian modeling, namely, Bayesian modeling that accounts for the two-component nature of plasmas, consisting of ions and electrons.

This importance is not limited to CTS analysis, but is also relevant to data-integration problems in which different diagnostics or data sets have different parameter sensitivities and their information is integrated through a common posterior distribution. In this sense, Bayesian inference provides a systematic framework for extracting physical parameters from multiple complementary observations while preserving high-dimensional correlations among model parameters. This perspective can be extended to the integrated analysis of multiple plasma diagnostics and, more generally, to inverse problems in which different data sets constrain different parts of a high-dimensional parameter space.

It should be noted that the EPW spectrum considered here has additional peaks that cannot be explained by a Maxwellian velocity distribution function \cite{zhang23nphys,ji24pop}. Non-Maxwellian velocity distribution functions are sometimes required in CTS analysis when spectra cannot be explained by a Maxwellian model \cite{morita22pre} and when \AN{observed signals are} clearly \AN{distinct from} those predicted by an equilibrium model \cite{sakai20pop,sakai23pop}. To further investigate the details of the velocity distribution function, non-Maxwellian models should be incorporated into Bayesian inference \cite{sakai26pop}.

\appendix
\section{Combined analysis without separate normalization}
\label{app:all_region}


\AN{The combined analysis described in Sec.~\ref{sec:3.3_combine} normalizes the IAW and EPW spectra independently so that neither spectral feature dominates the residual function simply because of its signal magnitude or wavelength resolution. To clarify the role of this separate normalization, we also examined a combined analysis without separating the IAW and EPW features. This corresponds to applying the uniform normalization condition to the entire wavelength range rather than separately to the IAW and EPW features. The calculated and experimental spectra are then normalized as}

\begin{figure}[!t]
    \centering

    \includegraphics[width=0.45\textwidth]{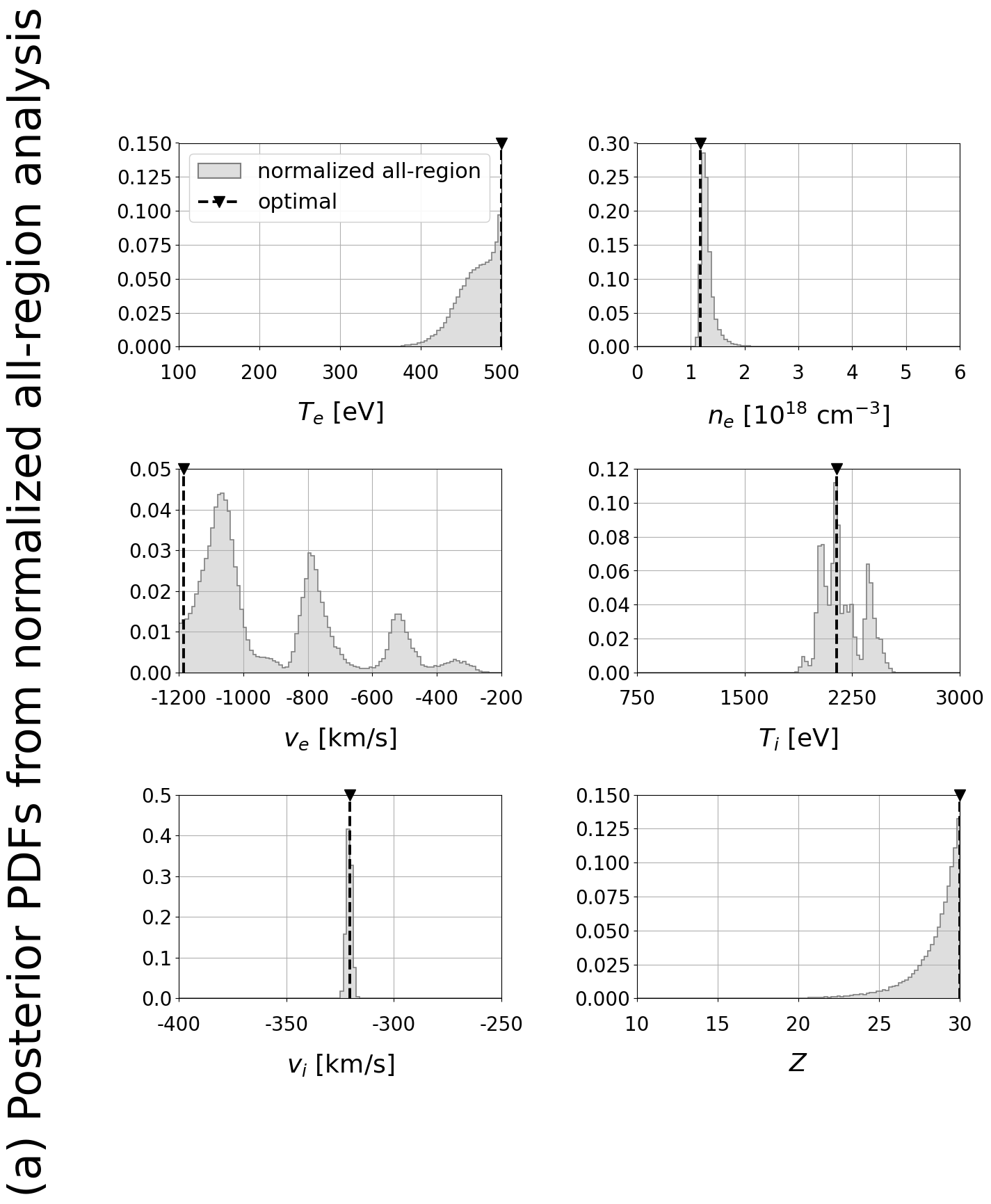}

    \vspace{2mm}

    \includegraphics[width=0.29\textwidth]{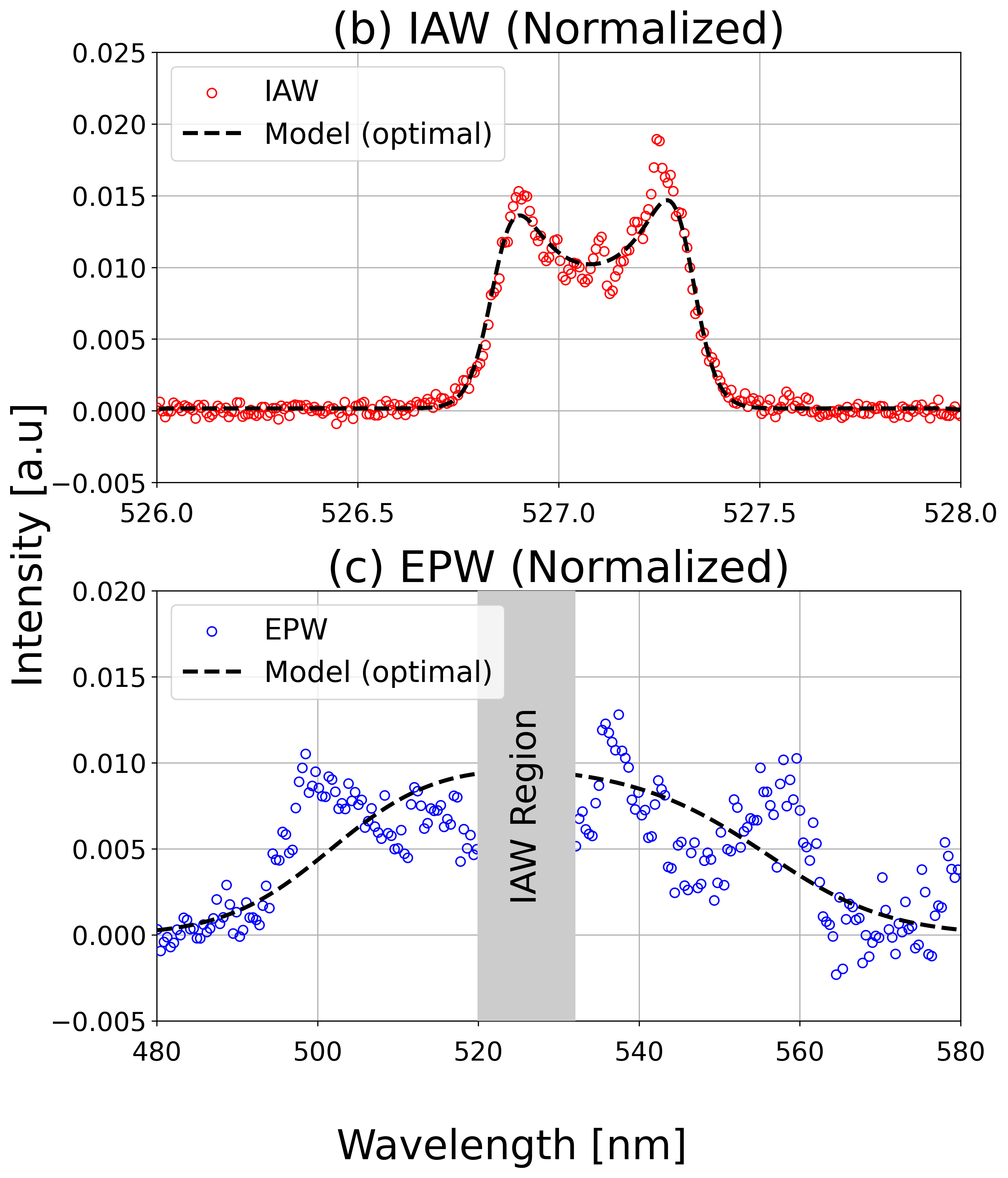}

    \caption{\AN{Result of the combined analysis without separate normalization at $\beta = 2.33 \times 10^5$: (a) 1D marginal posterior PDFs, (b) IAW spectrum, and (c) EPW spectrum.}}
    \label{fig:comb_all}
\end{figure}
\begin{eqnarray}
\sum_{\lambda_i \in \mathcal{R}_{\rm all}} I_{\rm cal}(\lambda_i;X)
=
\sum_{\lambda_i \in \mathcal{R}_{\rm all}} I_{\rm exp}(\lambda_i)
= 1,
\end{eqnarray}
and the residual function is defined as
\begin{eqnarray}
F_{\rm all}(X;Y)
=
\sum_{\lambda_i \in \mathcal{R}_{\rm all}}
\left[
I_{\rm cal}(\lambda_i;X)
-
I_{\rm exp}(\lambda_i)
\right]^2 .
\end{eqnarray}
Here, $\mathcal{R}_{\rm all}$ denotes the full wavelength region including both the IAW and EPW features.

\AN{Figure~\ref{fig:comb_all} shows the same type of plot as Fig.~\ref{fig:combine}, but for the result obtained using the all-region normalization.
With this normalization, the IAW data are not distinguished from the EPW data, despite their different wavelength resolutions and signal intensities.}
As a result, the inference is dominated by the IAW feature, reflecting its stronger signal intensity, and the EPW feature is not reproduced satisfactorily.
The resulting \AN{1D marginal posterior PDFs} are similar to those obtained from the IAW-only analysis in Fig.~\ref{fig:IAW}, indicating that the EPW data are effectively neglected.
This comparison supports the use of the separate normalization introduced in Sec.~\ref{sec:3.3_combine}.

\FloatBarrier
\vspace{-2mm}

\section{Combined analysis with data-point weighting}
\label{app:data_point_weighting}

\rev{The combined analysis described in Sec.~\ref{sec:3.3_combine} assigns equal weights to the separately normalized IAW and EPW residuals. To assess the sensitivity to this weighting choice, we additionally performed an analysis in which the residuals are weighted according to the respective numbers of wavelength data points. The data-point-weighted residual function is defined as}

\begin{flalign}
&F_{\rm weighted}(X;Y) \nonumber \\
&= \frac{1}{\left(N_{\rm IAW}+N_{\rm EPW}\right)}
\left[
N_{\rm IAW}F_{\rm IAW}(X;Y)
+ N_{\rm EPW}F_{\rm EPW}(X;Y)
\right]. &&
\label{eq:data_point_weighted}
\end{flalign}

\begin{figure}[!t]
    \centering

    \includegraphics[width=0.45\textwidth]{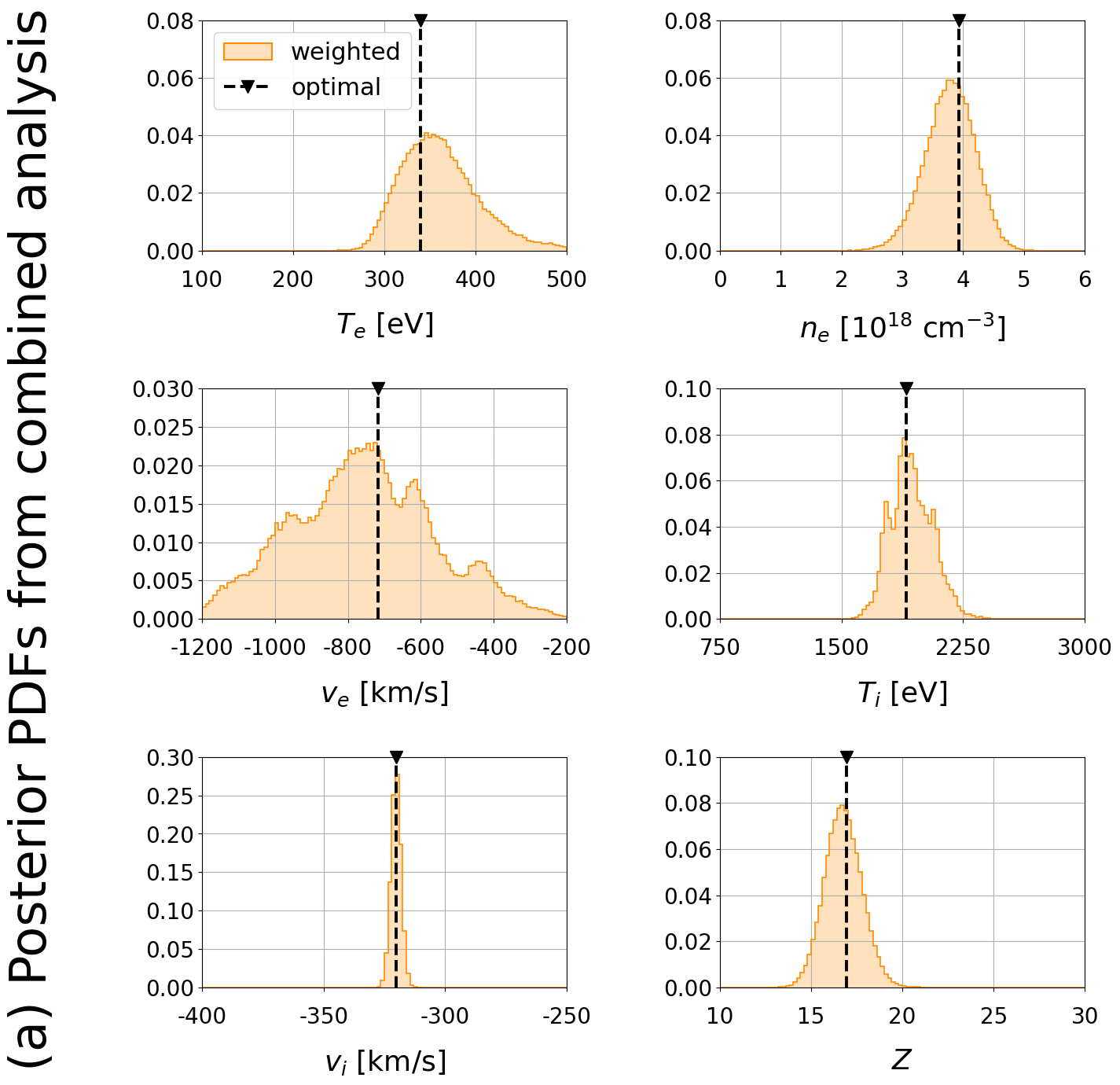}

    \vspace{2mm}

    \includegraphics[width=0.29\textwidth]{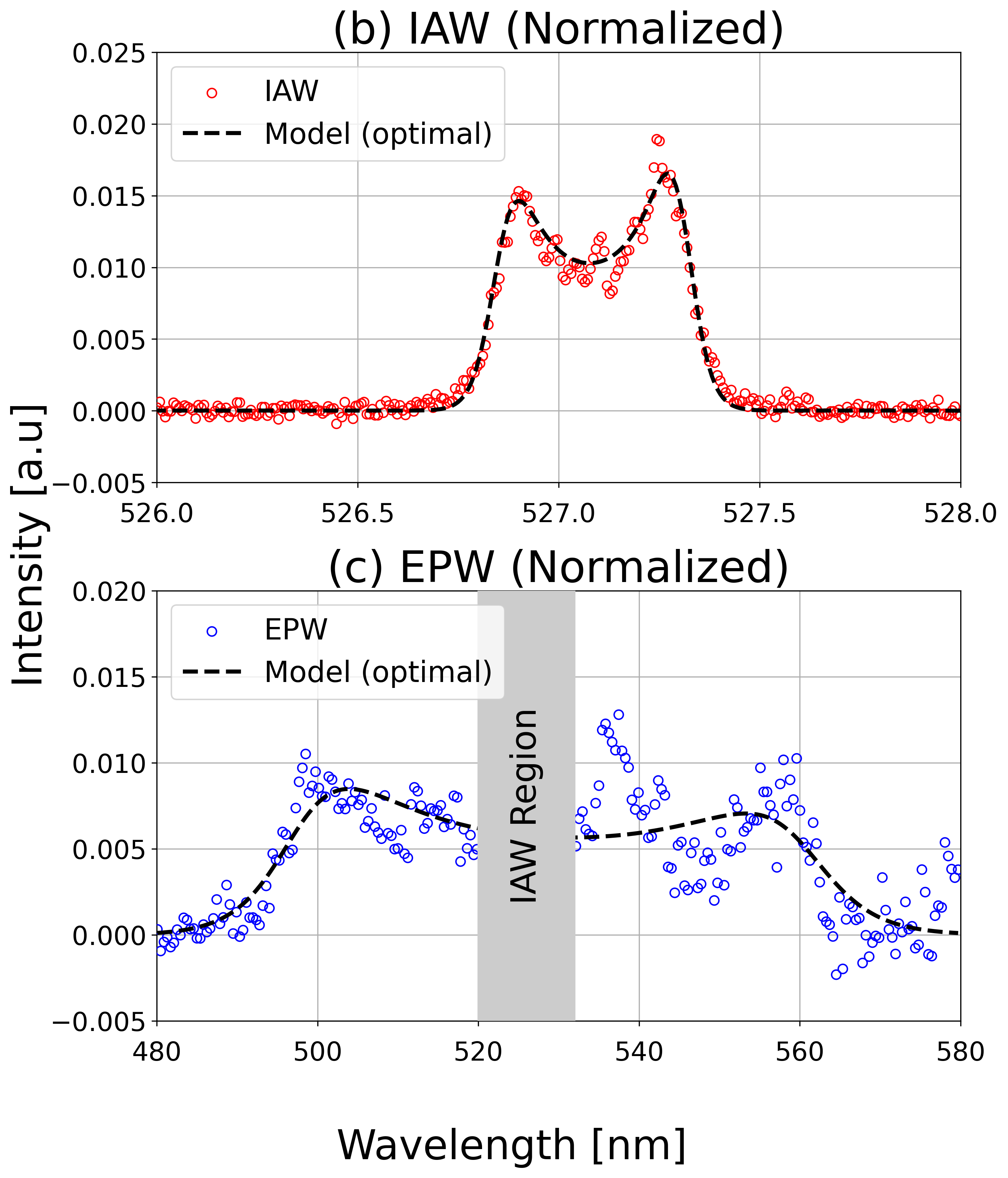}

    \caption{\rev{Result of the combined analysis with data-point weighting at $\beta = 2.33 \times 10^5$: (a) 1D marginal posterior PDFs, (b) IAW spectrum, and (c) EPW spectrum.}}
    \label{fig:weighted}
\end{figure}

\rev{Here, $N_{\rm IAW}$ and $N_{\rm EPW}$ denote the numbers of wavelength data points used in the IAW and EPW analyses, respectively. In the present data set, $N_{\rm IAW}=292$ and $N_{\rm EPW}=215$. Because each spectrum is normalized separately so that its discrete intensity sum is unity, the factors $N_{\rm IAW}$ and $N_{\rm EPW}$ compensate for the difference in residual scale associated with the spectral sampling density. Equation~\eqref{eq:data_point_weighted} therefore weights the two residuals according to their numbers of data points, whereas Eq.~\eqref{equ:combined} assigns equal weights. When $N_{\rm IAW}=N_{\rm EPW}$, Eq.~\eqref{eq:data_point_weighted} reduces to Eq.~\eqref{equ:combined}.}

\rev{Figure~\ref{fig:weighted} shows the result of the data-point-weighted analysis. Although the detailed shapes of the posterior distributions are modified, the overall parameter sensitivities remain similar to those obtained with equal weighting. In particular, the ion charge state $Z$ remains clearly constrained, indicating that the main conclusion of the present study is robust against this weighting choice. Weighting based explicitly on experimental uncertainties or S/N ratios would require a more detailed statistical model of the different noise characteristics of the IAW and EPW measurements and is left for future work.}

\section*{Acknowledgements}

A.N., K.S., and T.H. were supported by the JST Moonshot R\&D Program, Japan (Grant Number JPMJMS24A3), for the data analysis using ODAT-SE.
The program code was developed and the computations were performed using the facilities of the Supercomputer Center, the Institute for Solid State Physics, the University of Tokyo (ISSPkyodo-SC-2022-Ca-0037, 2023-Ca-0122, 2024-Ca-0120, 2025-Ca-0130, 2026-Ca-0155), on the supercomputer Fugaku through the HPCI System Research Project (Project ID: hp220248, hp230304, hp240304, hp250295), and on the supercomputer Wisteria-O through the Joint Usage/Research Center for Interdisciplinary Large-scale Information Infrastructures (JHPCN) in Japan (Project ID: jh240039, jh250034). \rev{T. H.  was supported for the code development partially by JSPS KAKENHI (Grant Numbers JP25H00449 and JP25K15153).}
K.S. acknowledges support from the Tatematsu Foundation of Public Interest. 
H.J., L.G., and B.K.R. were supported by the High Energy Density Laboratory Plasma Science program by the DOE Office of Science and NNSA under Grant No. DE-SC0020103. 

\bibliographystyle{unsrt}
\bibliography{ref}

\end{document}